\documentclass[notitlepage,showpacs,nobalancelastpage,aps,prl,twocolumn]{revtex4-2}
\usepackage[english]{babel}
\RequirePackage[T1]{fontenc}
\RequirePackage{times}

\usepackage{siunitx}[=v2] 
\usepackage[italicdiff]{physics}
\usepackage{amsfonts}
\usepackage{subfigure}
\usepackage{amsmath}
\usepackage{amssymb} 
\usepackage{graphicx,epstopdf}
\usepackage{appendix}
\usepackage{xspace}
\usepackage{lipsum}
\usepackage{array}
\usepackage[hidelinks]{hyperref}
\usepackage[dvipsnames]{xcolor}
\usepackage{my_symbols}
\usepackage{CJK}
\usepackage{bm}
\usepackage{verbatim}
\usepackage{tikz}
\usepackage{booktabs}
\usepackage{multirow}
\usepackage{pifont}
\usepackage[normalem]{ulem}
\usepackage{threeparttable}
\usepackage{setspace}

\begin{document}
\title{Spin wave vortex as topological probe of magnetic texture}
\author{Hongbin Wu}
\affiliation{Center for Joint Quantum Studies and Department of Physics, School of Science, Tianjin University, 92 Weijin Road, Tianjin 300072, China}
\affiliation{Tianjin Key Laboratory of Low Dimensional Materials Physics and Preparing Technology, Tianjin University, Tianjin 300354, China}
\author{Jin Lan}
\email[Corresponding author:~]{lanjin@tju.edu.cn}
\affiliation{Center for Joint Quantum Studies and Department of Physics, School of Science, Tianjin University, 92 Weijin Road, Tianjin 300072, China}
\affiliation{Tianjin Key Laboratory of Low Dimensional Materials Physics and Preparing Technology, Tianjin University, Tianjin 300354, China}

\begin{abstract}
A vortex, a circulating flow around a void, is one of the basic topological phenomena in nature. 
Here we show that vortices generally emerge in spin wave travelling upon topologically nontrivial magnetic texture, due to the transverse precession of spin wave about the background magnetization.
The winding number of each spin wave vortex is equivalent to sign of the local topological density of magnetic texture at the vortex core, and all winding numbers add up as twice the topological number of the magnetic texture. 
Based on the charts of spin wave vortices, the magnetization profile of the magnetic texture is reversely constructed, and a universal relation for the magnon topological Hall angle is theoretically proposed and numerically confirmed in vast types of magnetic textures.
The simple connection between dynamic and static magnetizations, promotes spin wave vortex as a powerful tool to reveal the topology of the underlying magnetic texture.
\end{abstract}

\maketitle

\emph{Introduction.}
Vortex is a fundamental and ubiquitous concept in both classical and quantum branches of physics, which spans from  galaxy in cosmology, typhoon in airflow, whirlpool in fluids \cite{saffman_vortex_1993,wu_vorticity_2006,donnelly_quantized_1991,abo-shaeer_observation_2001,freilich_realtime_2010,aharon-steinberg_direct_2022}, fluxon in superconductors \cite{abrikosov_magnetic_1957,blatter_vortices_1994} to structured field in waves \cite{gbur_singular_2016,forbes_structured_2021,jukes_formation_2013,zhang_creation_2020,zou_orbital_2020}. 
Depending on the timescale, vortices generally divide into two categories, the static configuration that hosts extra energy, and the dynamic structure that harbors additional information.  
Representatively, the static vortex is a key ingredient in type II superconductivity \cite{bardeen_theory_1965,tinkham_introduction_2004} and the  Berezinskii-Kosterlitz-Thouless phase transition \cite{nagaosa_quantum_1999,jose_40_2013}, and dynamic vortex is the primary carrier of orbital angular momentum in  optical \cite{coullet_optical_1989,shen_optical_2019} and electron beam \cite{uchida_generation_2010,mcmorran_electron_2011}.

Benefiting from the diverse timescales of magnetizations, the static and dynamic phenomena coexist in the same magnetic system \cite{gurevich_magnetization_1990}. 
As spatial and temporal excitations in magnets, the magnetic texture and spin wave, together with their rich interaction scenarios, attract paramount interest for both scientific explorations and industrial applications \cite{lan_spinwave_2015,han_mutual_2019,yu_magnetic_2020,yu_magnetic_2021,mahmoud_introduction_2020}.
The magnetic vortex \cite{shinjo_magnetic_2000,wachowiak_direct_2002}, along with magnetic domain wall \cite{parkin_magnetic_2008,caretta_relativistic_2020}, skyrmion \cite{yu_realspace_2010,fert_skyrmions_2013,jiang_skyrmions_2017}, bimeron \cite{gobel_magnetic_2019} and hopfion \cite{wang_currentdriven_2019,liu_threedimensional_2020}, serve as the basic storage units in racetrack memories.
In contrast, the spin wave vortex is still a missing member in the vortex family, and its interplay with magnetic texture remains elusive.

In this work, we show that vortices naturally develop in spin wave, and capture the essential topology of its phase structure. 
The winding numbers of the spin wave vortices are intimately connected to the topological number of magnetic texture, due to the topological connections between dynamic and static magnetizations.
From the charts of spin wave vortices, the texture magnetization is reconstructed, and the magnon topological Hall angles at vast types of magnetic textures are evaluated in a unified way.

\begin{figure*}[t]
	\centering
     {\includegraphics[width=0.99\textwidth,trim=5 0 0 0,clip]{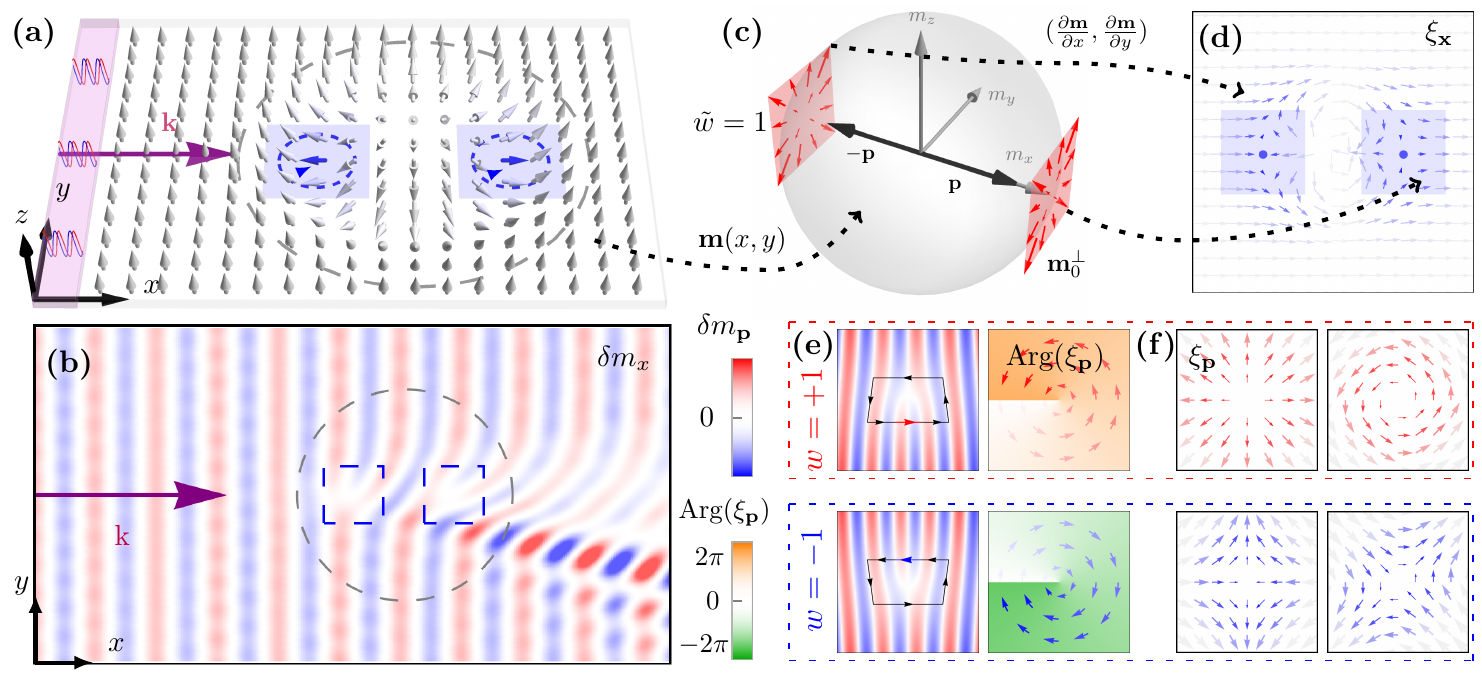}}
	\caption{{\bf Spin wave vortices on a topologically nontrivial magnetic texture.}
	(a) Schematics of a topologically nontrivial magnetic texture. The texture magnetizations are represented by gray arrows, and the spin wave antenna is represented by the purple bar.
	(b) Spin wave profile $\delta m_x$ extracted from micromagnetic simulations. The magnetic setup follows the schematics in (a), and the local regions around singular points are highlighted by dashed rectangles. 
	(c) Schematics of a magnetic Bloch sphere and its two tangent planes. The transverse magnetizations projected into the tangent planes  are  depicted by blue arrows.
	(d) Vectorial representation of the projected basis $\xi_\bp$ around two singular points $\mb_0=\pm \bp$.
	(e) Typical profiles  and the corresponding phase structure of spin wave vortices for positive/negative winding number $w=\pm 1$.  
	(f) Typical profiles of the projected basis $\xi_\bp$ for $w=\pm 1$. 
	}
	\label{fig:sw_vortex}
\end{figure*}

\emph{Singular points in spin wave profile.}
Consider a ferromagnet film in $x$-$y$ plane with its magnetization direction denoted by unit vector $\mb$, as depicted in Fig. \ref{fig:sw_vortex}(a).
The magnetization naturally partitions into the static magnetic texture part $\mb_0$ and the dynamic spin wave part $\delta\mb$, $\mb=\mb_0+\delta \mb$. 
In the small amplitude limit of spin wave $|\delta \mb|\ll 1$, the transverse condition is satisfied everywhere $\delta \mb\cdot \mb_0=0$, as enforced by the unity constraint $|\mb|=|\mb_0|=1$.
Hence, by attaching spherical coordinates $\hbe_{r/\theta/\phi}$ to background texture magnetization $\hbe_r\equiv \mb_0$, the spin wave is described by $\delta \mb=\delta m_\theta\hbe_\theta+\delta m_\phi \hbe_\phi$. 
Furthermore, the spin wave can be always regarded as right circular about the static magnetization in uniaxial ferromagnet \cite{kim_landaulifshitz_2015,ai_anatomy_2023}, and thus is rewritten as $\delta \mb\equiv\Re[\psi \bm{\xi}]$, where $\psi=\delta m_\theta-i\delta m_\phi$ and $\bm{\xi}=\hbe_\theta+i \hbe_\phi$ are the complex wave component and basis  \cite{schutte_inertia_2014,liu_geometric_2023}, respectively.

Due to its vectorial nature, the spin wave is typically projected into a specific direction for experimental observation and theoretical inspection \cite{stancil_spin_2009,wintz_magnetic_2016,bertelli_magnetic_2020,qin_lowloss_2022} (e.g., $\delta m_x$ along the $\hbx$ direction). 
Generally, along arbitrary direction $\bp$, the projected spin wave component is $\delta m_\bp = \delta \mb\cdot \bp$, which follows relation $\delta m_\bp=\Re[\psi \xi_\bp]$ with $\xi_\bp=\bm{\xi}\cdot \bp$ the projected basis. 
Consequently, at some specific sites where the local texture magnetization aligns with the projection direction, $\mb_0=\pm \bp$,  the projected basis becomes exactly zero $\xi_\bp=0$.
The vanishing intensity (amplitude) of the spin wave component $\delta m_\bp$, along with the indetermined phase  \cite{dennis_topological_2001,dennis_chapter_2009,gbur_singular_2016,berry_singularities_2023}, then lead to singular points in the spin wave profile.

\emph{Spin wave vortex around a singular point}.
For convenience of inspection, we define a fixed frame $\hbe_{1/2/3}$, which coincides with the co-rotating frame $\hbe_{r/\theta/\phi}$ at a chosen singular point. 
At vicinity of the singular point, the projected basis reads 
\begin{align}
	\label{eqn:basis_ep}
	\xi_\bp \approx  m_2 + i  m_3,
\end{align}
where $m_{2/3}=\mb_0\cdot \hbe_{2/3}$ are static magnetization components in the transverse direction of $\hbe_1$. 
In above approximation, the relations $\hbe_\theta \cdot \hbe_1\approx \hbe_2\cdot \hbe_r$ and $\hbe_\phi \cdot \hbe_1\approx \hbe_3\cdot \hbe_r$ are invoked for these two neighboring frames.
Around the singular point, the projected basis $\xi_\bp$ expands as
\begin{align}
	\label{eqn:basis_expansion}
	\xi_\bp \approx
	 \begin{pmatrix} 1 & i \end{pmatrix} \cJ \begin{pmatrix} \Delta x \\ \Delta y  \end{pmatrix}, 
\end{align}
where  $\cJ$ is the Jacobian matrix given by 
\begin{align}
	\label{eqn:jacobian}
	\cJ = \begin{pmatrix} \pdv{m_2}{x} & \pdv{m_2}{y} \\
		 \pdv{m_3}{x} &  \pdv{m_3}{y}  \end{pmatrix},
\end{align}
and $(\Delta x,\Delta y)$ are the displacements from the singular point.

Without loss of generality, consider a planar spin wave $\psi = c e^{i kx }$, where the complex amplitude $c$ and the wave vector $k$ are both regarded as constants.
Modulated by the projected basis $\xi_\bp$ with the typical profiles shown in Fig. \ref{fig:sw_vortex}(f), the spin wave $\delta m_\bp$ forms a vortex around the singular point in Fig. \ref{fig:sw_vortex}(e).
The topology of vortex is characterized by the winding number
\begin{align} 
	\label{eqn:winding_number}
	w\equiv \frac{1}{2\pi} \oint_l d\arg(\psi \xi_\bp) \equiv \frac{1}{2\pi} \oint_l d\arg(\xi_\bp)= \mathrm{sgn}(J),
\end{align}
where $l$ is a small loop enclosing the singular point, and $J$ is the Jacobian $J\equiv \det (\cJ)$.
Excluding the laminar flow, the vorticity of the spin wave flux circulating around the vortex core is $|c|^2J \hbz$.
As a defining feature of a vortex,  the spin wave experiences a wavefront dislocation \cite{nye_dislocations_1974,nye_natural_1999,dutreix_measuring_2019,dutreix_wavefront_2021,liu_visualizing_2024} with an additional half-wavefront developed at the singular point, as illustrated in Fig. \ref{fig:sw_vortex}(e). 
For vortices with winding number $w=\pm 1$, an additional phase of $\pm 2\pi$ is accumulated along the loop $l$, and an extra half-wavefront is inserted from the lower/upper side of the spin wave fringe.

As a representative example, consider that a planar spin wave travels upon a magnetic skyrmion, as schematically depicted in Fig. \ref{fig:sw_vortex}(a). 
For projection direction $\bp=\hbx$, there are two points with local magnetization aligns with magnetization $\mb_0=\pm \hbx$. 
In Fig. \ref{fig:sw_vortex}(d), the projected basis $\xi_\bx$ becomes zero at these two singular points, and forms two saddle-type structures at the vicinity with winding number $w=-1$. 
Consequently,  two vortices are unambiguously identified by two fork patterns in the fringe, at the spin wave profile $\delta m_x$ extracted from micromagnetic simulation in Fig. \ref{fig:sw_vortex}(b).
Moreover, due to the insertion of two additional half-wavefronts from the upper side, the fringe is squeezed to the lower side, giving rise to an intensified beam toward the lower-right direction.

\emph{Spin wave vortex and magnetic texture topology.}
The texture magnetization profile $\mb_0(x,y)$ can be deemed as a map from the complex plane $x+iy$ to a magnetic Bloch sphere in Fig. \ref{fig:sw_vortex}(c), where the all peripheral region collapse to a single $\infty$ point \cite{dubrovin_modern_1985}. 
As a map between two compact manifolds of the same dimension $CP^1\to S^2$, the degree of map $Q$ (more known as topological number \cite{nagaosa_topological_2013,gobel_skyrmions_2021}) is an integer, which counts the number of times the complex plane wraps the  Bloch sphere under the mapping $\mb_0(x,y)$. 
In integral form, the topological number (degree of map) is written  as
\begin{align}
\label{eqn:Q_int}
Q = \iint q dx dy=\iint \qty[\frac{1}{4\pi} \mb_0 \cdot \qty(\pdv{\mb_0}{x}\times \pdv{\mb_0}{y})]dx dy,
\end{align}
where $q$ is the topological density (Gaussian curvature normalized by $4\pi$ \cite{dubrovin_modern_1985}).

For an arbitrary direction $\bp$, there are two points lying oppositely on the Bloch sphere $\mb_0=\pm\bp$, with the preimage points in the $x$-$y$ plane as the singular points. 
The local mapping from the region around a singular point in $x$-$y$ plane to the tangent plane in the Bloch sphere is governed by the Jacobian matrix $\cJ$ in \Eq{eqn:jacobian} \cite{dubrovin_modern_1985}. 
Moreover, the sign of the Jacobian $\mathrm{sgn}(J)$ determines the relative (same or opposite) orientation between the $x$-$y$ plane and the tangent plane, and thus relates the original winding number $w$ and the target winding number  $\tilde{w}$ via $w=\mathrm{sgn}(J)\tilde{w}$. 
It is noteworthy that the Jacobian in \Eq{eqn:winding_number} is related to topological density in \Eq{eqn:Q_int} by $J =4\pi q$, and the transverse magnetization in the tangent plane $\mb_0^\perp=\bp\times (\mb_0\times \bp)$ is the vectorial form of  $\xi_\bp$ in \Eq{eqn:basis_ep}. 
Hence, observing that the winding number of the transverse magnetization is fixed to $\tilde{w}=1$ in Fig. \ref{fig:sw_vortex}(c), the winding number of the spin wave vortex in the $x$-$y$ plane is then given by \Eq{eqn:winding_number}.

\begin{figure}[tb]
	\centering
     {\includegraphics[width=0.5\textwidth,trim=5 0 0 0,clip]{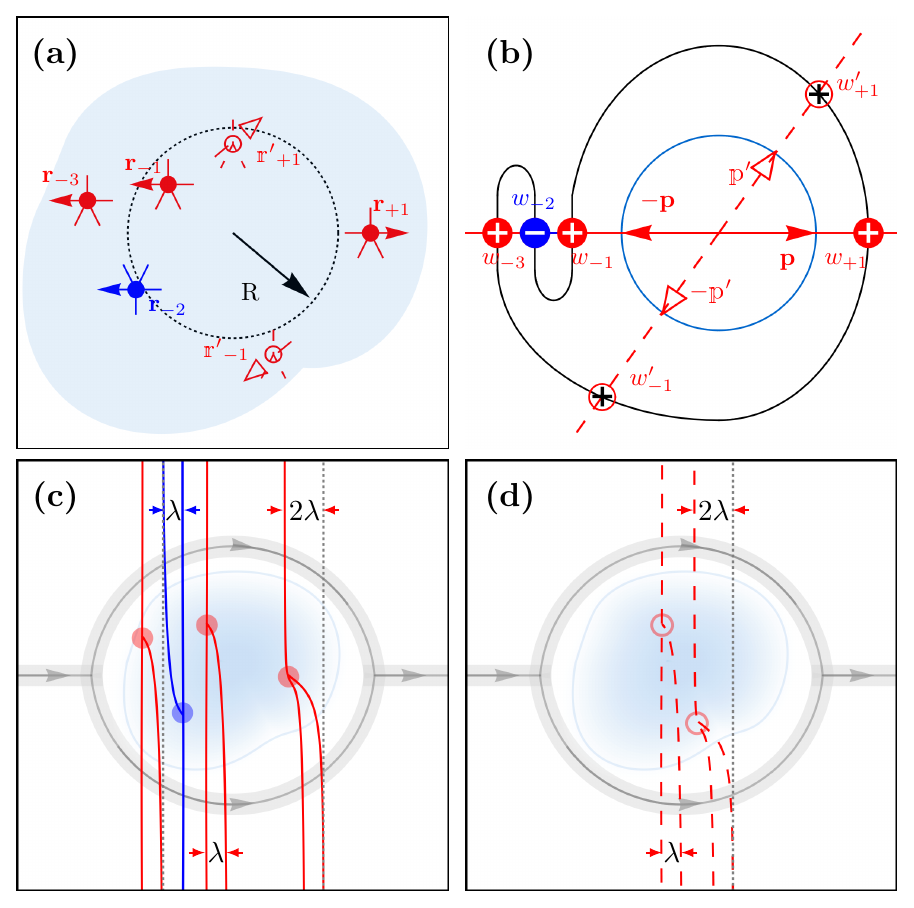}}
	\caption{
	{\bf Distributions of spin wave vortices and the corresponding wavefront dislocations.}
    (a) Distributions of spin wave vortices in real space. 
	(b) Distributions of singular points in mapped spin space.
	Positive/negative signed circles indicate the inner/outer side of the map pierced by the radial lines emanating from origin.
	(c)(d) The wavefront dislocations along $\bp$ and $\bp'$. 
	Only the main wavefronts in fork shape are sketched, and the background color encode the topological density $q$. 
	In all figures, the solid/dashed (or filled/hollow) elements are along the observation directions $\bp$ and $\bp'$, respectively. }
	\label{fig:map_2D}
\end{figure}

\begin{figure}[tb]
	\centering
    {\includegraphics[width=0.5\textwidth]{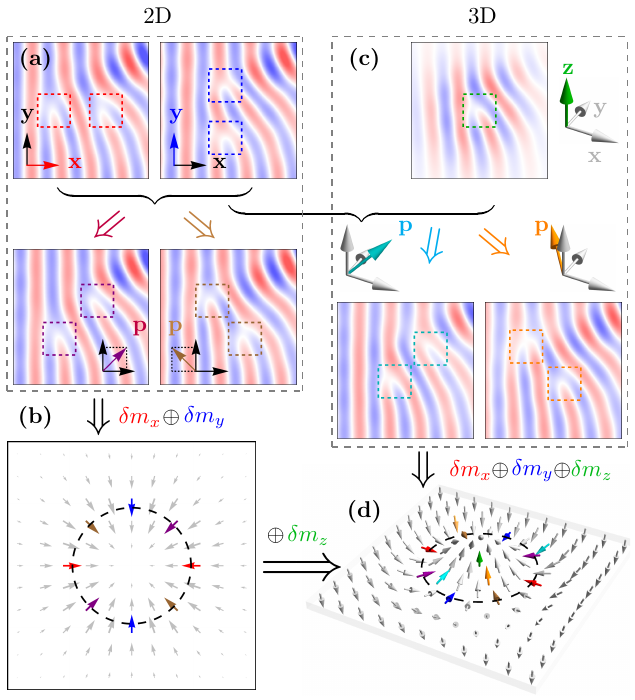}}
	\caption{{\bf Reconstruction of texture magnetization from spin wave vortices.   } 
	(a) Upper panels: Original spin wave  projected into $\hbx$ and $\hby$ directions; Lower panels: Reconstructed spin wave projected into in-plane directions $(\pm\hbx+ \hby)/\sqrt{2}$.
	(b) Deduced texture magnetizations along in-plane directions in $2$-dimensions.
	(c) Upper panel: Original spin wave projected into $\hbz$ directions; Lower panels: Reconstructed spin wave projected into directions $(\pm\hbx+ \hby+\hbz)/\sqrt{3}$.
	(d) Deduced texture magnetization along selected directions in full $3$-dimensions.
	}
	\label{fig:skyrmion_combine}
\end{figure}

\emph{Global Aharonov-Bohm phase and local phase singularities.}
The spin wave dynamics in uniaxial ferromagnets is generally governed by a Schrodinger-like equation \cite{volovik_linear_1987,lan_spinwave_2015,kim_tunable_2019,lan_skew_2021}
\begin{align}
	i \partial_t\psi  =\gamma A \qty[(i\nabla-\ba)^2 ]\psi,
\end{align}
where $\gamma$ is the gyromagnetic ratio, $A$ is the exchange  coupling constant.
Here, $\ba$ is the fictitious vector potential mediated by the magnetic topology \cite{vanhoogdalem_frequencydependent_2012,tatara_effective_2019,kim_tunable_2019,lan_skew_2021}, and $b =(\nabla\times \ba)_z=4\pi q$ is the fictitious magnetic field \cite{iwasaki_theory_2014,schutte_magnonskyrmion_2014,lan_geometric_2021,liang_bidirectional_2023}. 
The Aharonov-Bohm phase acquired by a spin wave along the loop enclosing the whole magnetic texture is given by \cite{berry_wavefront_1980,gorodetski_plasmonic_2010, cohen_geometric_2019,iwasaki_theory_2014,schutte_magnonskyrmion_2014}
\begin{align}
	\label{eqn:AB_phase}
	\Delta \Phi= \oint \ba \cdot d\br = \iint b   dxdy = 4\pi Q, 
\end{align}
which is quantized in $2\pi$ to ensure the single-valuedness of spin wave.

Meanwhile, since the indeterminacy of phase is excluded in regions with finite wave amplitude, a non-zero phase only accumulates along the loops encircling spin wave vortices. 
Hence, the Aharonov-Bohm phase in \Eq{eqn:AB_phase} divides into phase singularities in spin wave vortices,
\begin{align}
	\label{eqn:phase_sum}
	 \Delta\Phi= 2\pi \qty(\sum_{j}w_{+j}+\sum_{j'}w_{-j'} ), 
\end{align}
where $w_j$ ($w_{-j'}$) is the winding number of  spin wave vortex at $\br_{+j}$ ($\br_{-j}$) with $\mb_0=\pm \bp$. 
Alternatively, $\Delta\Phi/2\pi$ measures the relative shift between spin wave fringes in homogeneous domains far above and below the magnetic texture, by counting the number difference of half-wavefronts inserted from upper and lower sides, as illustrated in Fig. \ref{fig:map_2D}(c)(d).

As two aspects of the same phase, the equivalence between \Eq{eqn:AB_phase} and \Eq{eqn:phase_sum} are actually ensured by the identity \cite{dubrovin_modern_1985}
\begin{align}
	\label{eqn:q_w}
	 Q= \sum_{j}w_{+j} =\sum_{j'}w_{-j'}, 
\end{align} 
which is the discrete form of the topological number (degree of map).
As schematically depicted in Fig. \ref{fig:map_2D}(b), rather than integrating $q$ throughout the whole map in \Eq{eqn:Q_int}, a more straighforward evaluation of $Q$ is to count the in/out sides of the map pierced by a ray in arbitrary direction $\bp$, characterized by $w_j=\mathrm{sgn}[q(\br_j)]$ \cite{wang_equivalent_2010,sticlet_geometrical_2012,asboth_short_2016}. 
According to \Eq{eqn:q_w}, the spin wave vortices always emerge in pair, grouped in co-/counter-rotating fashion with the same/opposite winding number.
Particularly for $|Q|=1$, there are at least one co-rotating pair of spin wave vortices within the magnetic texture, as a manifestation of the Hopf-Poincare index (hairy ball) theorem \cite{needham_visual_2021}.
As the observation direction $\bp$ varies, the positions and windings of spin wave vortices change correspondingly in Fig. \ref{fig:map_2D}(a), but their combinations are always subject to the constraints of the topological number $Q$.

\emph{Texture magnetization from spin wave vortices.} 
Based on three independent profiles along orthogonal directions $\qty{\hbx_{1/2/3}}$, the spin wave profile along arbitrary direction $\bp$ can be generally synthesized, $\delta\mb \cdot\bp= \sum_{j=1}^{3} (\delta\mb \cdot \hbx_j)(\hbx_j\cdot\bp)$. 
By treating the vortices as position indicators in the synthesized spin wave profile along $\bp$, the texture magnetization $\mb_0=\pm \bp$ is reversely identified in the magnetic film. 
Collecting the $2$D anatomy slices of spin wave profiles by sweeping $\bp$, the $3$D full figure of texture magnetization is reconstructed, as illustrated in Fig. \ref{fig:skyrmion_combine}(c)(d).

Even when certain spin wave information is inevitably missing in practice, the texture magnetization can still be  partially retrieved.
In Fig. \ref{fig:skyrmion_combine}(a)(b), without out-of-plane spin wave component $\delta m_z$, the spin wave profile along arbitrary in-plane direction is available from $\delta m_x$ and $\delta m_y$, and all texture magnetizations residing in in-plane directions are identified in the magnetic film.

\begin{figure}[t!]
	\centering
	{\includegraphics[width=0.5 \textwidth]{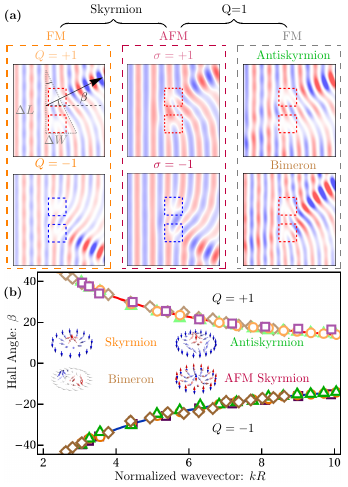}}
    \caption{
		{\bf Spin wave vortices (a) and magnon topological Hall angle (b) for vast types of magnetic textures with $|Q|=1$.}
        In (a), the background red/blue color encodes the spin wave profile, the dashed squares highlight the locations of spin wave vortices. 
		In (b),  the dots are  extracted from micromagnetic simulations, and the solid lines represent the universal relation in \Eq{eqn:hall_angle}.
		\label{fig:hall_angle}
        }
\end{figure}

\emph{Magnon topological Hall angle.}
In a wave perspective, the change of propagation direction is dictated by the transverse shift $\Delta W$ and longitudinal shift $\Delta L$ associated with the wavefront bending, as illustrated in Fig. \ref{fig:hall_angle}(a).
Translating each $2\pi$ phase into a wavelength $\lambda$, the transverse shift is $\Delta W = 2\lambda Q $ according to \Eq{eqn:AB_phase}.
Meanwhile, the longitudinal shift is formulated by $\Delta L=2\eta R$, where $R$ is the characteristic size of the magnetic texture, and $\eta$ is the shape factor that phenomenologically characterizing spatial distributions of spin wave vortices.
With the geometry built upon above two perpendicular shifts, the magnon toplogical Hall angle is given by
\begin{align}
	\label{eqn:hall_angle}
	\beta = \arctan \frac{\Delta W}{\Delta L}
	= \arctan \qty(\frac{\pi Q }{\eta}\frac{1}{kR}),
\end{align}
which reduces to the inverse proportionality $\beta\propto 1/kR$ in the short wavelength limit $k R\gg 1$.
Except the topological number $Q$ \cite{nagaosa_topological_2013,schutte_magnonskyrmion_2014,iwasaki_theory_2014} and normalized wavevector $kR$ \cite{iwasaki_theory_2014}, the Hall angle $\beta$ in  \Eq{eqn:hall_angle} is solely determined by the shape factor $\eta$ of the magnetic texture.
Although the shape factor $\eta$ varies per definition, the Hall angle $\beta$ is only tunable via changing the wavevector $k$ of spin wave  or the size $R$ of magnetic texture, once a specific type of texture is chosen. 

The universal Hall angle $\beta$ in \Eq{eqn:hall_angle} is verified by micromagnetic simulations using Mumax 3 \cite{vansteenkiste_design_2014} for vast types of magnetic texture, as plotted in Fig. \ref{fig:hall_angle}(b). 
Particularly for three types of magnetic texture with $|Q|=1$,  magnetic skyrmion, bimeron and anti-skyrmion, the Hall angle $\beta$ as function of normalized wavevector $kR$ falls into the same line with shape factor $\eta\approx 2.5$, for a wide range of magnetic texture size $R$ and spin wave wavevector $k$. 
As anchors for the shape factor, the co-rotating vortice pair always reside on a circle of radius $R$ within all these textures, when the observation direction is perpendicular to the magnetization in homogeneous domain.
As the topological number $Q$ increases to $Q=2$ and further $Q=3$, the Hall angle increases only slightly, due to low efficiency in assembling spin wave vortices to accommodate the final wavefront bending \cite{SM}.

In antiferromagnets and ferrimagnets, the texture magnetizations and the spin wave polarizations hosted by two magnetic sublattices are opposite \cite{kittel_introduction_1953,cheng_spin_2014,lan_antiferromagnetic_2017,rezende_introduction_2019}. 
Seeing opposite sublattice magnetizations and topologies, the left-/right-circular spin waves develop oppositely winding vortices across the same antiferromagnetic texure, and thus exhibit polarization-dependent Hall angles \cite{daniels_topological_2019,kim_tunable_2019} in Fig. \ref{fig:hall_angle}.

\emph{Wave and particle perspectives.}
Focusing on the intensity profile or phase structure of wave field, the particle/wave perspective lies at two opposite extremes in disentangling intensity and phase degrees of freedom. 
As two complementary approaches, both perspectives are able to probe topological defects: the former hinges on trajectory bending of wave packet under emergent electromagnetic fields \cite{xiao_berry_2010,bliokh_geometrodynamics_2009,nagaosa_emergent_2012,lan_skew_2021,tatara_effective_2019}, while the latter relies on the conversion from plane wave to vortex under interference \cite{masajada_optical_2001,wang_recent_2018,wang_photonic_2019,dutreix_measuring_2019,dutreix_wavefront_2021,zhang_local_2020,liu_visualizing_2024}.
Counting discrete phase jumps around singular points, instead of scrutinizing continuous intensity distortions in extended areas, the vortex provides a more compact and comprehensive toolkit to analyze the wave scattering upon a given environment.

\emph{Conclusion.}
In conclusion, we demonstrate that spin wave vortices are ubiquitous upon topologically nontrivial magnetic textures, due to the intimate topological connection between dynamic and static magnetizations.
With wavefront geometry built upon charts of spin wave vortices, a universal relation for magnon topological Hall angle is revealed.
The deep interweaving between topology and phase, boosts vortex as a vigorous turbo in exploring and analyzing the wave field.


%

\end{document}